\documentclass[a4paper,12pt]{article}
\usepackage{graphicx,rotating,hyperref,slashed,amsmath,xcolor,amssymb,amsfonts,cite}
\pdfoutput=1
\makeatletter
\hypersetup{colorlinks,bookmarksopen,bookmarksnumbered,
linkcolor=blus,pdfstartview=FitH,urlcolor=rossos,citecolor=verde}

\usepackage{tikzfeynman}

\def\lsim{\mathrel{\rlap{\lower3pt\hbox{\hskip0pt$\sim$}}
   \raise1pt\hbox{$<$}}}         
\def\gsim{\mathrel{\rlap{\lower4pt\hbox{\hskip1pt$\sim$}}
   \raise1pt\hbox{$>$}}}         

\newcommand{\mio}[1]{}

\newcommand{\fig}[1]{~\ref{fig:#1}}

\definecolor{Gray}{gray}{0.95}

\renewcommand{\L}{{\cal L}}
\newcommand{\qL}{\theta_L}
\newcommand{\qD}{\theta_D}
\newcommand{\qR}{\theta_R}

\newcommand{\Q}{{\cal Q}}

\newcommand{\pb}{\,{\rm pb}}
\newcommand{\fb}{\,{\rm fb}}

\usepackage{multicol}
\usepackage{color}
\definecolor{rosso}{cmyk}{0,1,1,0.4}
\definecolor{rossos}{cmyk}{0,1,1,0.55}
\definecolor{rossoc}{cmyk}{0,1,1,0.2}
\definecolor{blu}{cmyk}{1,1,0,0.3}
\definecolor{blus}{cmyk}{1,1,0,0.6}
\definecolor{bluc}{cmyk}{1,1,0,0.1}
\definecolor{verde}{cmyk}{0.92,0,0.59,0.25}
\definecolor{verdec}{cmyk}{0.92,0,0.59,0.15}
\definecolor{verdes}{cmyk}{0.92,0,0.59,0.4}

\newcommand{\riga}[1]{\noalign{\hbox{\parbox{\textwidth}{#1}}}\nonumber}

\newcommand{\eq}[1]{~{\rm (\ref{eq:#1})}}

\newcommand{\GeV}{\,{\rm GeV}}
\newcommand{\TeV}{\,{\rm TeV}}

\def\circa#1{\,\raise.3ex\hbox{$#1$\kern-.75em\lower1ex\hbox{$\sim$}}\,}

\newcommand{\beq}{\begin{equation}}
\newcommand{\eeq}{\end{equation}}

\newcommand{\bea}{\begin{eqnarray}}
\newcommand{\eea}{\end{eqnarray}}
\newcommand{\be}{\begin{equation}}
\newcommand{\ee}{\end{equation}}
\font\tenrsfs=rsfs10 at 12pt
\font\sevenrsfs=rsfs7 at 10 pt
\font\fiversfs=rsfs5
\newfam\rsfsfam
\textfont\rsfsfam=\tenrsfs
\scriptfont\rsfsfam=\sevenrsfs
\scriptscriptfont\rsfsfam=\fiversfs
\def\mathscr#1{{\fam\rsfsfam\relax#1}}

\def\Lag{\mathscr{L}}

\def\circa#1{\,\raise.3ex\hbox{$#1$\kern-.75em\lower1ex\hbox{$\sim$}}\,}
\makeatletter

\def\hhref#1{\href{http://arxiv.org/abs/#1}{arXiv:#1}} 

\def\hhref#1{\href{http://arxiv.org/abs/#1}{arXiv:#1}} 
 
\def\art{\@ifnextchar[{\eart}{\oart}}
\def\eart[#1]#2#3#4#5#6{{\rm #2}, {\em #3 \bf #4} {\rm (#6) #5} ({\em #1})}

\def\article{\@ifnextchar[{\earticle}{\oarticle}}
\def\oarticle#1#2#3#4#5#6{{\rm #1}, {\em ``#6''}, {\rm #2 #3 (#5) #4}}
\def\earticle[#1]#2#3#4#5#6#7{{\rm #2}, {\em ``#7''}, {\rm #3 #4 (#6) #5}  [\hhref{#1}]}
\def\hepart[#1]#2{{\rm #2, \em#1}}
\def\heparticle[#1]#2#3{#2, {\em ``#3''} [\hhref{#1}]}

\newcounter{alphaequation}[equation]
\def\thealphaequation{\theequation\hbox to
0.6em{\hfil\alph{alphaequation}\hfil}}
\def\eqnsystem#1{
\def\@eqnnum{{\rm (\thealphaequation)}}
\def\@@eqncr{\let\@tempa\relax \ifcase\@eqcnt \def\@tempa{& & &} \or
  \def\@tempa{& &}\or \def\@tempa{&}\fi\@tempa
  \if@eqnsw\@eqnnum\refstepcounter{alphaequation}\fi
\global\@eqnswtrue\global\@eqcnt=0\cr}
\refstepcounter{equation} \let\@currentlabel\theequation \def\@tempb{#1}
\ifx\@tempb\empty\else\label{#1}\fi
\refstepcounter{alphaequation}
\let\@currentlabel\thealphaequation
\global\@eqnswtrue\global\@eqcnt=0 \tabskip\@centering\let\\=\@eqncr
$$\halign to \displaywidth\bgroup \@eqnsel\hskip\@centering
$\displaystyle\tabskip\z@{##}$&\global\@eqcnt\@ne
\hskip2\arraycolsep\hfil${##}$\hfil& \global\@eqcnt\tw@\hskip2\arraycolsep
$\displaystyle\tabskip\z@{##}$\hfil
\tabskip\@centering&\llap{##}\tabskip\z@\cr}
\def\endeqnsystem{\@@eqncr\egroup$$\global\@ignoretrue} \makeatother

\newcommand{\SU}{\,{\rm SU}}

\newcommand{\U}{\,{\rm U}}
\begin{document}

\centerline{CERN-PH-TH-2015-311\hfill IFUP-TH/2015}
\bigskip
\bigskip

\begin{center}
{\LARGE \bf \color{rossos} Trinification can explain the\\[2mm] 
di-photon and di-boson LHC anomalies}\\[1cm]

\bigskip

{\large\bf 
Giulio Maria Pelaggi$^a$,
Alessandro Strumia$^{a,b}$,
Elena Vigiani$^a$
}  
\\[5mm]

\bigskip

{\it $^a$ Dipartimento di Fisica dell'Universit{\`a} di Pisa and INFN, Italy}\\[1mm]
{\it $^b$ CERN, Theory Division, Geneva, Switzerland}


\vspace{1cm}
{\large\bf\color{blus} Abstract}
\begin{quote}\large
LHC data show a diphoton excess at 750 GeV and a less significant diboson excess  around 1.9 TeV.
We propose trinification as a common source of both anomalies.
The 1.9 TeV excess can be produced by the lightest extra vector: a $W_R^\pm$  with a gauge coupling $g_R\approx 0.44$ 
that does not decay into leptons.
Furthermore, trinification predicts extra scalars. One of them
can reproduce the $\gamma\gamma$ excess 
while satisfying constraints from all other channels,
given the specific set of extra fermions predicted by trinification.
\end{quote}

\thispagestyle{empty}
\end{center}

\setcounter{page}{1}
\setcounter{footnote}{0}

\tableofcontents

\newpage

\section{Introduction}
After the discovery of the Standard Model Higgs boson~\cite{h}  at the Large Hadron Collider,
data indicate the presence of other excesses that can be interpreted as resonant production of new bosons.
The two most notable anomalies found in present LHC data are:
\begin{enumerate}
\item An excess of di-boson events peaked around $1.9\TeV$, found in run 1 data~\cite{pierini,data}.
This has been explained in terms of resonant production of extra vectors, such as $Z'$ or $W_R$
that decay into $ZZ$ or $WZ$, respectively~\cite{dibosonWZ}.
Run 2 data do not confirm this excess, but (as we will see) do not significantly contradict it.

\item An excess of di-gamma events peaked around $750\GeV$, mostly seen at run 2~\cite{data}.
This has been explained in terms of resonant production of an extra scalar $S$~\cite{gammagamma,goldrush}.
In order to achieve the needed  $S\to \gamma\gamma$ decay width, extra vector-like fermions or scalars coupled to $S$ 
with a relatively large coupling must be introduced~\cite{gammagamma,goldrush}.
\end{enumerate}
Assuming that these excesses are not statistical fluctuations, 
we try to go beyond phenomenological considerations and explore which kind of bigger picture they can suggest.

\smallskip

The structure needed to explain both anomalies
is present in models that extend the SM gauge group around the weak scale,
thereby implying extra vectors, predicting a larger set of anomaly-free 
chiral fermions and needing extra Higgs scalars and Yukawa couplings.
Among this class of models, trinification is the
$\SU(3)_L\otimes\SU(3)_R\otimes\SU(3)_c$ sub-group of E$_6$ that explains 
the observed quantised hypercharges,
that can be broken to the Standard Model
at a scale $V$ just above the weak scale, that allows a totally asymptotically free extension of the SM~\cite{333,333a},
which is suggested by non conventional ideas about naturalness of the Higgs mass~\cite{TAF}.
Trinification contains left-right models and 3-3-1 models as subgroups~\cite{331}.
Focusing  on trinification:
\begin{itemize}
\item It predicts extra $W_R^\pm$ vectors with mass $\sim g_RV$
with properties compatible with the di-boson anomaly: no decays into leptons, appropriate value of its $g_R$ gauge coupling.

\item It predicts 27 chiral fermions per generation: the SM fermions, plus
extra vector-like $D'$ and $L'$ fermions (and singlets), which receive masses  $M' \sim y' V$.
Large Yukawa couplings $y'$ to fermions are needed in order to make $M'$ above present bounds,
if $V$ is not much above the weak scale.

\item It predicts a larger set of Higgs scalars, which contain various singlets and doublets (with neutral components).
One of them can be identified with the Higgs doublet at 125 GeV, and another with the scalar $S$ at 750 GeV.

\end{itemize}
In section~\ref{trini} we briefly summarise the main ingredients of trinification models needed to interpret the LHC anomalies.
In section~\ref{diboson} we discuss how trinification can reproduce the di-boson anomaly.
In section~\ref{digamma} we discuss how trinification can reproduce the di-gamma anomaly.
Conclusions are given in section~\ref{concl}.

\begin{table}
$$
\begin{array}{|rccc|ccc|}\hline
\multicolumn{2}{|c}{\hbox{Field}}&\hbox{spin} &\hbox{generations} 
& \SU(3)_L&\SU(3)_R & \SU(3)_{\rm c}
\cr \hline
\Q_R\,=\!\!&{ \begin{pmatrix} U  &  D_1  &    D_2 \end{pmatrix}}^T&1/2&3& 1 & 3  & \bar 3 \cr
\Q_L\,=\!\!&{ \begin{pmatrix} Q & \bar D' \end{pmatrix}} &1/2&3& \bar 3 &1 &   3  \cr
\L\,=\!\!&{ \begin{pmatrix} \bar L'\ & L_2 & L_1 \cr E & N_2 & N_1 \end{pmatrix}} & 1/2& 3&3 & \bar 3  & 1  \cr
{\cal H} =\!\!& { \begin{pmatrix} H_u^* & H_d & H_L \cr S^+ & S_1& S_2 \end{pmatrix}}  &0&3  &3 & \bar 3  & 1  \cr 
 \hline
\end{array}$$
\caption{\em\label{tab:333} Field content of minimal weak-scale trinification.
$Q$ is the SM left-handed quark doublet,
$U$ is the SM right-handed up quark,
$E$ is the SM right-handed lepton.
The $D_1$ and $D_2$ quarks mix making $D$ (right-handed down quarks) and $D'$ (that gets a mass with $\bar D'$)
the $L_1$ and $L_2$ lepton doublets mix making $L$ (left-handed lepton doublet) and $L'$ (that gets a mass with $\bar L'$).
$N_1$ and $N_2$ are singlets.}
\end{table}

\begin{figure}[t]
\centering
$$\includegraphics[width=\textwidth]{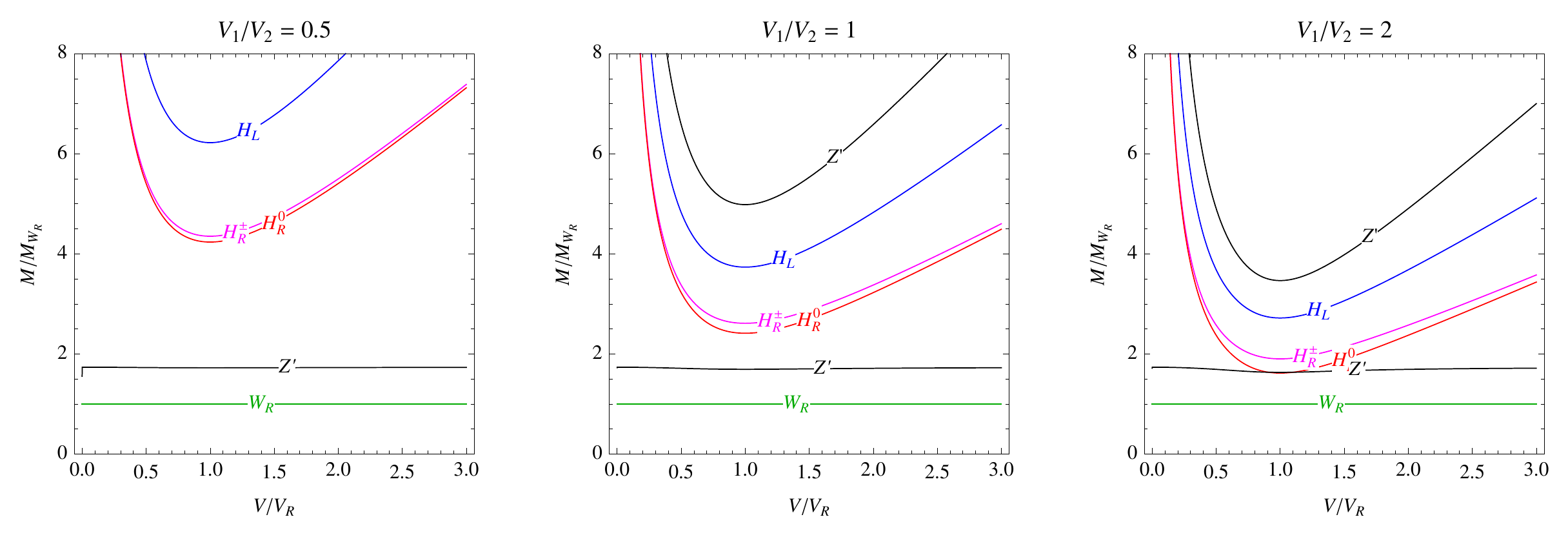}$$
\caption{\em Trinification heavy vector masses normalised to the $W_R$ mass. 
The next-to-lightest vector is a $Z'$ with mass $\approx 1.7 \, M_{W_R}$. 
For concreteness, we used a basis where $V_{R1}=0$.
\label{fig:MV}}
\end{figure}

\section{Trinification}\label{trini}
The minimal model of weak-scale trinification is obtained 
including 3 families of chiral-free fermions under the gauge group
$G_{333}=\SU(3)_L\otimes\SU(3)_R\otimes\SU(3)_c$
and 3 families of Higgses, as summarised in table~\ref{tab:333}.
The three trinification gauge coupling constants ($g_2$, $g_R$, $g_3$) allow to reproduce those of the SM ($g_2$, $g_Y$, $g_3$) as
\beq  
g_R = \frac{2 g_2 g_Y}{\sqrt{3 g_2^2 - g_Y^2}}  \approx 0.444.\eeq
The SM Yukawa couplings are obtained from the $G_{333}$-invariant interactions
\beq
- \Lag_{Y} = y_{Qn}^{ij} ~\Q_{Li} \Q_{Rj}  {\cal H}_n + \frac{y^{ij}_{Ln}}{2} {\cal L}_i{\cal L}_j {\cal H}_n +{\rm h.c.}
\label{eqqqY}
\eeq
The most generic vacuum expectation values that give the desired pattern of symmetry breaking  are
\beq 
\langle {\cal H}_n\rangle =  \begin{pmatrix} v_{un} & 0&  0 \\   0 &  v_{dn} &  v_{L n}\\  0 & V_{Rn} & V_n\end{pmatrix}.\label{eq:vevs}
\eeq
Defining 
$V^2\equiv \sum_n (V_n^2+V_{Rn}^2)$
and the dimension-less ratios 
\beq 
\alpha\equiv \sum_n V_{Rn}^2/V^2,\qquad
\beta\equiv \sum_n V_n V_{Rn}/V^2 \,,
\eeq 
in the relevant limit $v\ll V$ one obtains the following extra massive vectors:\footnote{Notice that trinification includes both left-right models (the group
$ \SU(2)_L\otimes\SU(2)_R \otimes{\rm U}(1)_{B-L}\otimes \SU(3)_c$
is obtained in the limit $\alpha,\beta\ll1$)
and 3-3-1 models~\cite{331} (the group $\SU(3)_L\otimes\SU(3)_c\otimes\U(1)_X$ is obtained
from trinification if $\SU(3)_R$ is broken to its $\U(1)$ component that commutes with $\SU(2)_R$.
This can be achived adding an extra Higgs scalar in the adjoint).
3-3-1 models have the same minimal set of chiral fermions as trinification, but do not have the $W_R$ vector that
can fit the di-boson anomaly.
Left-right models do not imply extra fermions, needed to fit the digamma anomaly.}
\begin{itemize}
\item A left-handed weak doublet with  mass
$M_{H_L} = g_2 V/\sqrt{2}$,
that contains two neutral components and two charged components.

\item Two $Z', Z''$ vectors with masses:
\beq M_{Z', Z''}^2 = \frac{V^2}{3} \bigg[ (g_2^2 + g_R^2)  \pm \sqrt{ (g_2^2 + g_R^2)^2 +3
g_R^2 (4g_2^2+g_R^2) (\alpha^2-\alpha+\beta^2 )} \bigg].\eeq

\item A $\SU(2)_R$ vector doublet $A_{2R}$ splitted into two neutral components with mass
$M_{H_R^0}= g_R V/\sqrt{2}$ and into 2 charged components with mass
\beq  M_{H_R^\pm}^2 = \frac{g_R^2 V^2}{4} \bigg[ 1 + \sqrt{(1-2\alpha)^2+4\beta^2}\bigg]  . \eeq
 
\item A $\SU(2)_R$ triplet $A_{3R}$ that splits into a neutral $Z'$ and into 
2 charged right-handed  $W_R^\pm$ vectors with mass
\beq  M_{W_R^\pm}^2 = \frac{g_R^2 V^2}{4} \bigg[ 1 -\sqrt{(1-2\alpha)^2+4\beta^2}\bigg] \,.  \eeq
\end{itemize} 
We see that $W_R$ is the lightest extra vector.
The $W_R^\pm$ and $H^\pm_R$ mass eigenstates arise as
\beq \begin{pmatrix} W_R^\pm  \cr H_R^{\pm}\end{pmatrix} =
\begin{pmatrix} \cos\qR & -\sin\qR\cr \sin\qR & \cos\qR\end{pmatrix} 
\begin{pmatrix} A_{R3}^\pm \cr A_{R2}^\pm \end{pmatrix}
\eeq
in terms of the $G_{333}$ interaction eigenstates $A_{R3}^\pm$ and $A_{R2}^\pm$. 
The mixing angle vanishes if $\beta\ll1$:
\beq \tan2\qR = \frac{2\beta}{1-2\alpha}.
\eeq
Concerning fermions, trinification adds to the chiral fermions under the SM group
($Q$, $D$, $U$, $L$ and $E$ in the usual notation), an extra vector-like $D'\oplus\bar D'$ and $L'\oplus \bar L'$
as well as some singlets $N$, $N'$.
In presence of three Higgs multiplets ${\cal H}$, the theory has enough Yukawa couplings that
the extra states can be naturally heavy, $M_{D',L'} \sim y_{Q,L} V$ thanks to 
large enough Yukawa couplings $y_{Q,L}$~\cite{333}.
The light and heavy fermion mass eigenstates can be parametrised as
\beq \begin{pmatrix} D \cr D'\end{pmatrix} =
\begin{pmatrix} \cos\qD & -\sin\qD\cr \sin\qD & \cos\qD\end{pmatrix} 
\begin{pmatrix} D_1 \cr D_2\end{pmatrix},
\qquad
 \begin{pmatrix} L \cr L'\end{pmatrix} =
\begin{pmatrix} \cos\qL & -\sin\qL\cr \sin\qL & \cos\qL\end{pmatrix} 
\begin{pmatrix} L_1 \cr L_2\end{pmatrix}
  \eeq
where $\qD$ and $\qL$ are (for each generation) 
mixing angles, that can be of order one, computable in terms of Yukawa couplings and vevs.
Then, the $W_R$ coupling to quark mass eigenstates is 
\beq 
\frac{1}{\sqrt{2}} \, (\tilde{g}_R \bar{d}_R+\tilde{g}'_R \bar{d}'_R)\,  \slashed{W}_{R}^+ u_R + \hbox{h.c.} \qquad
\tilde{g}_R= g_R\cos(\qD +\qR),\qquad
\tilde{g}'_R = g_R \sin(\qD+\qR) \,.
\eeq
An important feature predicted by trinification is that
the $W_R$ does not couple to two light leptons.
Rather, $W_R$ couples the SM doublet of left-handed leptons $L$ to the heavy doublet of anti-leptons $\bar L'$
with coupling $g_R\sin(\theta_R-\theta_L)$, 
and right-handed leptons $E$ to heavy neutrinos with coupling $g_R \cos\theta_R$, in the limit of negligible mixing angle
between neutral fermions $N_1/N_2$.
Furthermore, $W_R$ couples to pairs of heavy leptons.

\begin{figure}[t]
\centering
\begin{center}
\begin{tikzpicture}[line width=1 pt, scale=1.55]
 \draw[fermion] (0.1,0.5)--(0.6,0);
        \draw[fermion] (0.6,0)--(0.1,-0.5);
   	\draw[fermion] (1.9,0.5)--(1.4,0);
   	\draw[fermion] (1.4,0)--(1.9,-0.5);
        \draw[vector] (0.6,0)--(1.4,0);
   	\node at (-0.1,0.7) {$q_R$};
   	\node at (2.1,0.7) {$\bar{q}_R '$};
    \node at (-0.1,-0.7) {$\bar{q}_R '$};
    \node at (2.1,-0.7) {$q_R$};
    \node at (1,-0.3) {$W_R^{\pm}$};
        \draw[fermion] (3.3,0.5)--(3.8,0);
        \draw[fermion] (3.8,0)--(3.3,-0.5);
   	\draw[vector] (5.1,0.5)--(4.6,0);
   	\draw[vector] (5.1,-0.5)--(4.6,0);
        \draw[vector] (3.8,0)--(4.6,0);
   	\node at (3.1,0.7) {$q_R$};
   	\node at (5.3,0.7) {$W^{\pm}$};
    \node at (3.1,-0.7) {$\bar{q}_R '$};
    \node at (5.3,-0.7) {$Z$};
    \node at (4.2,-0.3) {$W_R^{\pm}$};
    \draw[fermion] (6.3,0.5)--(6.8,0);
        \draw[fermion] (6.8,0)--(6.3,-0.5);
   	\draw[vector] (8.1,0.5)--(7.6,0);
   	\draw[dashed] (8.1,-0.5)--(7.6,0);
        \draw[vector] (6.8,0)--(7.6,0);
   	\node at (6.1,0.7) {$q_R$};
   	\node at (8.3,0.7) {$W^{\pm}$};
    \node at (6.1,-0.7) {$\bar{q}_R '$};
    \node at (8.3,-0.7) {$h$};
    \node at (7.2,-0.3) {$W_R^{\pm}$};
\end{tikzpicture}
\end{center}
\caption{\em Resonant production of a  $W_R^\pm$ that decays into $jj$, $W Z$ and $W h$.
\label{fig:jj-WZ-Wh}}
\end{figure}
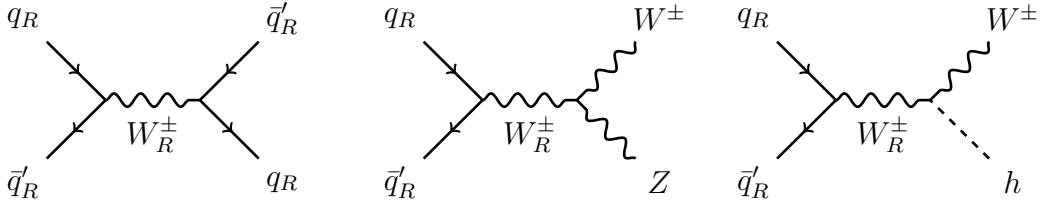




\begin{table}[t]
$$\begin{array}{cccccc}
\hbox{process} & \hbox{mass} & \multicolumn{2}{c}{\sigma  \hbox{ at }\sqrt{s}=8\TeV} &\multicolumn{2}{c}{\sigma  \hbox{ at }\sqrt{s}=13\TeV}\\ \hline
pp\to S\to \gamma\gamma &750\GeV &  0.5\pm0.5 \fb  &\hbox{\cite{data,CMSgg8,ATLASgg8}}&  8\pm 2\fb & \hbox{\cite{data}}\\
pp\to W_R^\pm \to  W^\pm Z & 1.9 \TeV & 5.3\pm 2.0\fb &\hbox{\cite{pierini}}& 1 \pm 13\fb &\hbox{\cite{data,ATLAS13-WZ,CMS13-WZ}}\\
pp\to W_R^\pm \to  W^\pm h & 1.9 \TeV &3.2\pm 3.7\fb& \hbox{\cite{ATLAS8-WH,CMS8-WH,CMS8-WH-lvbb}}&37\pm 32 \fb & \hbox{\cite{ATLAS13-WH}}\\
pp\to W_R^\pm \to  jj & 1.9 \TeV &73\pm39\fb &\hbox{\cite{jjbound,ATLAS8-jj,CMS8-jj}} & 53\pm 230\fb & \hbox{\cite{ATLAS13-jj,CMS13-jj}}\\
\end{array}$$
\caption{\em Observed cross sections for the anomalous digamma and diboson excesses at LHC that we try to interpret
in terms of trinification.\label{tabounds}
}
\end{table}

\section{The di-boson excess at 1.9 TeV}\label{diboson}
LHC data show an excess of di-boson events around 1.9 TeV~\cite{pierini}.
Trinification can explain this excess as $pp\to W_R^\pm\to W^\pm Z$,
and predicts an equal rate for $pp\to W_R^\pm\to W^\pm h$,
some amount of $pp\to W_R^\pm\to jj$ but no $pp\to W_R^\pm\to \ell^\pm \nu$, as discussed below.
No excess is observed in di-leptons, while $jj$ data show some excess.
The measured rates are summarised in table~\ref{tabounds}.
Fig.\fig{jj-WZ-Wh} shows the Feynman diagrams that describe such processes.

\subsection{Trinification predictions}

The cross section for producing a on-shell $W_R^\pm$ in narrow-width approximation is
\beq
\sigma(pp\to W_R^\pm ) = K \,\frac{\pi \tilde{g}_R^2}{6s} \, C_\pm(s),
\eeq
where the dimensionless partonic coefficients are 
\begin{eqnarray}
C_+(s)&=&\sum_{u,d}
|V_{ud}|^2
\int dy  \,
  f_{u} \left(\frac{M_{W_R}}{\sqrt s} e^y\right) f_{\bar d} \left(\frac{M_{W_R}}{\sqrt s} e^{-y}\right)   \,, \\
C_-(s)&=&\sum_{u,d}
|V_{ud}|^2
\int dy \,
  f_{\bar u} \left(\frac{M_{W_R}}{\sqrt s} e^y\right) f_{ d} \left(\frac{M_{W_R}}{\sqrt s} e^{-y}\right)   \,,  
\end{eqnarray}
to be integrated in the rapidity range $|y|< \ln\sqrt{s}/M_{W_R}$, summing over all up-type and down-type quarks.
The factor $K = 1 + {8 \pi} \alpha_3(M_{W_R}^2)/9\approx 1.23$ 
accounts for leading-order QCD corrections, with
the partonic function distributions $f_q$ renormalised at $Q=M_{W_R}$.
The numerical values of the $C_\pm$ coefficients are given in table~\ref{sigmaWR}
using the MSTW2008NLO partonic distribution functions~\cite{pdfs}.


\begin{table}[t]
\beq\begin{array}{c|cc}
& \sqrt{s}=8\TeV & \sqrt{s}=13\TeV\\  \hline
\sigma(pp\to W_R^\pm) & 0.75 \pb \, \times \, \tilde{g}_R^2/ \rho^{6.4} & 5.3 \pb \,\times  \,\tilde{g}_R^2 /\rho^{4.8} \\[2mm]
C_+ (s) & 0.14 /\rho^{6.4} & 2.6  /  \rho^{4.7} \\
C_-(s) & 0.047/ \rho^{6.5} & 0.92/ \rho^{5.1}\\
\end{array}\eeq
\caption{\label{sigmaWR}\em  Cross section $\sigma(pp\to W_R^\pm)$ and partonic factors $C_\pm$,
as function of the $W_R$ mass, assumed to be close to $1.9\TeV$.
We defined $\rho = M_{W_R}/1.9 \TeV$.}
\end{table}

\medskip

The cross sections into given final states $X$ are
\beq \sigma(pp\to W_R^\pm \to X) = \sigma(pp\to W_R^\pm ) \, \times \, {\rm BR}(W_R^\pm\to X).\eeq
We then need the decay widths and the branching ratios. 
Simple results hold into two extreme limits, described below.
\begin{enumerate}

\item All fermions and scalars are much lighter than $M_{W_R}$, so that the
$W_R$ decay width into full $G_{333}$ multiplets is simply given by
\beq
\label{Gamma_QQ}
\Gamma(W_R^\pm \to Q_R \bar Q_R) = \Gamma(W_R^\pm \to L \bar L) = 
2\Gamma(W_R^\pm \to {\cal H}^*{\cal H})=
\frac{g_R^2}{16 \pi}  M_{W_R} 
\eeq
independently of the mixing angle $\qR$ that defines $W_R$.
Taking into account that there are 3 generations of fermions and scalars, the total width is
$\Gamma_{W_R}/M_{W_R}=15 g_R^2 /32\pi \approx 0.029$.

\item $W_R$ can only decay into the SM fermions and into the Higgs doublet, while all
other states are so heavy that decays into them are kinematically forbidden. 
Interestingly, $W_R$ cannot decay into leptons.  The decay width into a light quark generation is
\begin{equation}
\label{Gamma_qq}
\Gamma(W_R^+ \to u \bar{d}) =  \frac{\tilde{g}_R^2 |V_{ud}|^2}{16 \pi} \, M_{W_R} \,.
\end{equation}

The $W_R$ decay width into the components of the SM Higgs doublet $H = (\chi^\pm, (h + i \eta)/\sqrt{2})$ 
(where, after $\SU(2)_L$ breaking, $\chi^\pm$ and $\eta$ become the longitudinal components of $W^\pm$ and $Z$, respectively) is 
\begin{equation}
\label{Gamma_WZ}
\Gamma(W_R^\pm \to W^\pm Z) =\Gamma(W_R^\pm \to W^\pm h)=   \frac{{g}_R^2 }{192 \pi} \, M_{W_R} \cos^2\theta_H\,\,,
\end{equation}
where $\theta_H$ is a mixing angle that defines which components of ${\cal H}$ contain the light SM Higgs doublet $H$,
given by $\cos\theta_H \equiv r_{ud} \cos \theta_R-r_{uL} \sin\theta_R$ with\footnote{The decay width into SM vector bosons can also be computed taking into account that
the small SM vacuum expectation values $v$ mixes $W_R$ with the SM $W$ by a small angle 
\beq
\theta_{LR} \simeq \frac{g_R}{g_2} \, \frac{M_W^2}{M_{W_R}^2} \, \cos\theta_H \,,\qquad
\eeq
and that
\begin{eqnarray}
\label{Gamma_WZ}
\Gamma(W_R^\pm \to W^\pm \, Z) =\Gamma(W_R^\pm \to W^\pm \, h) = \frac{g_2^2  \theta_{LR}^2}{192 \pi}  \frac{M_{W_R}^5}{M_W^4}
= \frac{g_R^2  }{192\pi } M_{W_R}\cos^2\theta_H.
\end{eqnarray}
}
\beq r_{ud} = \frac{2\sum_n v_{un} v_{dn}}{\sum_n (v_{un}^2+v_{dn}^2 + v_{Ln}^2)} \,,  \qquad
r_{uL} = \frac{2\sum_n v_{un} v_{Ln}}{\sum_n (v_{un}^2+v_{dn}^2 + v_{Ln}^2)} .\eeq
In this limit, the total decay width is $\Gamma_{W_R}/M_{W_R}=  g_R^2(18 \cos^2(\theta_D+\theta_R) + \cos^2\theta_H)/96\pi$.

\end{enumerate}
One loop QCD corrections enhance the decay widths into quarks by  $K_D= 1+ {\alpha_3(M_{W_R})}/{\pi} \approx 1.03$.

If all the mixing factors are of order unity, in  case 2 one expects $\hbox{BR}(W_R\to WZ)\sim 1/36$.
However this branching ratio can be much larger if $\cos^2\theta_H\gg \cos^2(\theta_D+\theta_R)$.
In case 1 one has $ \hbox{BR}(W_R\to WZ)=\hbox{BR}(W_R\to Wh)\le 1/90$.

\begin{figure}[t]
\begin{center}
$$\includegraphics[width=0.48\textwidth]{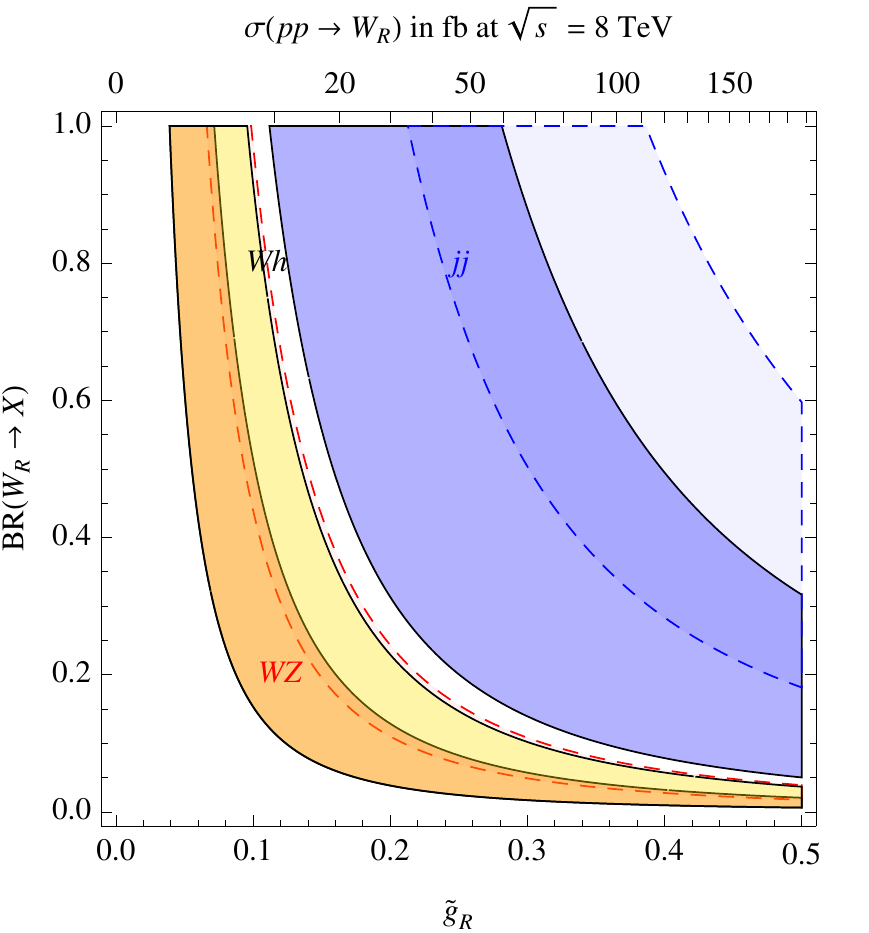}\quad
\includegraphics[width=0.45\textwidth]{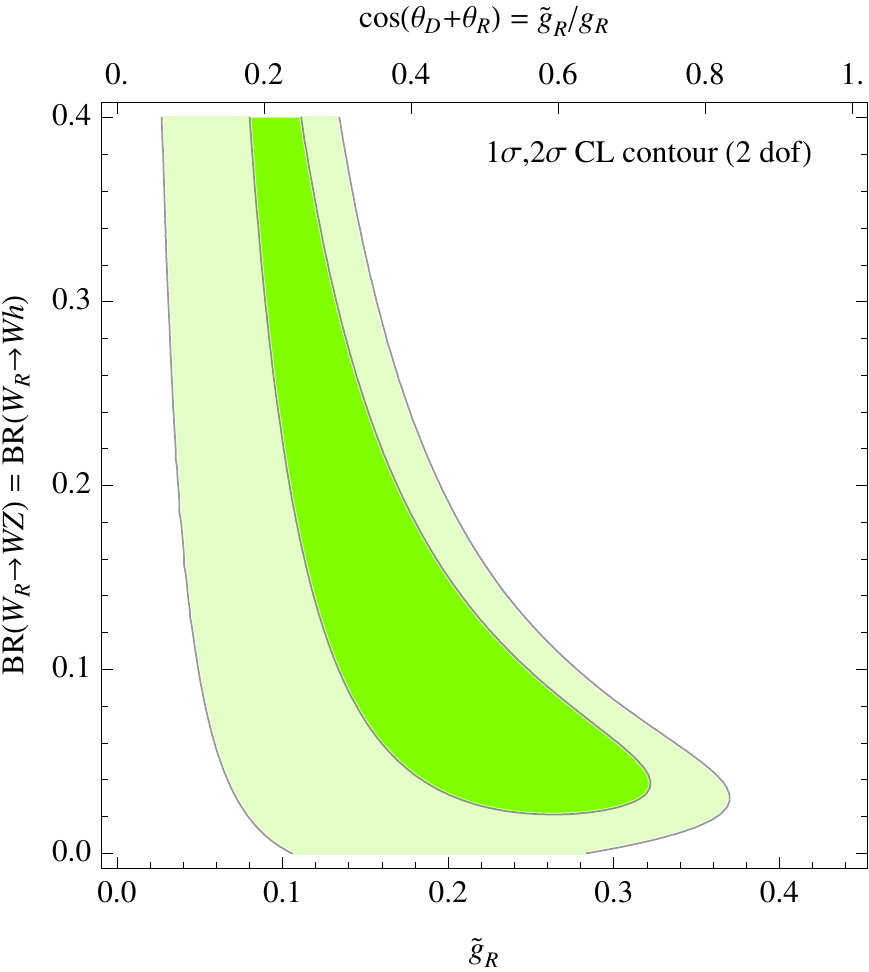}$$
\caption{\em \label{fig:fit} Global fit of run 1 and run 2 data.
{\bf Left}: $\pm1\sigma$ experimental bands for the $W_R$ Branching Ratios into
$jj$ (blue), $WZ$ (red), $Wh$ (yellow) 
as function of the coupling $\tilde g_R$ that controls the predicted 
$W_R$ production cross section.
We also show the bands obtained using $8\TeV$ data (dashed contours with lighter colors), and the
cross section at $8\TeV$ (upper axis).
{\bf Right}: global fit (green region) assuming only $W_R$ decays into $jj, WZ, Wh$.}
\end{center}
\end{figure}

\subsection{Comparison with data}
We are now ready to compare the trinification predictions with di-boson data.

The first issue that we address is the compatibility of run 1 data at $\sqrt{s}=8\TeV$
(that showed an excess around 1.9 TeV) with run 2 data at $\sqrt{s}=13\TeV$, that show no excess so far.
In view of the reduced luminosity so far accumulated at run 2, and of the larger backgrounds,
run 2 data, as summarised in table~\ref{tabounds}, have much larger uncertainties.
In our model the $W_R$ production cross section grows by a factor $\approx 7$
when going from $\sqrt{s}=8$ to 13 TeV.

Taking into account all these factors, the incompatibility between the two data-sets is only mild, at the $1\sigma$ level,
and the run 1 data still play the main role.
This is shown in the left panel of fig.\fig{fit}, where we extract the $W_R$ branching ratios into the various channels as function
of the $W_R$ production cross section, plotted at 8 TeV on the upper horizontal axis, and parametrised in terms of the coupling $\tilde g_R$
on the lower axis.

\medskip

We next proceed in performing a global fit, assuming that $W_R$ only decays into $jj, WZ$ and $Wh$ (case 2 in the discussion above)
and using the theoretical prediction $\hbox{BR}(W_R\to WZ) = \hbox{BR}(W_R\to Wh)$.
The result is shown in the right panel of fig.\fig{fit}.
We see that data prefer $\tilde{g}_R \sim 0.25$.
While in a generic context this is a free parameter, trinification
predicts  $\tilde g_R =g_R \cos(\qD+\qR)$,
where $g_R=0.444$ and the mixing factor depends on the details of the model.
This means that trinification is non-trivially compatible with data,
that favour a substantial amount of mixing among right-handed down quarks, $\cos(\qD+\qR)\sim 0.5$.
The best fit has $\chi^2_{\rm best} = \chi^2_{\rm SM} - 2.7^2$, which can be interpreted as a $2.7\sigma$ hint,
up to look-elsewhere reductions.
Similar results are obtained for a slightly lighter or heavier $W_R$.

Allowing $W_R$ to decay into extra light states (up to the opposite extremum of case 1 in the discussion above)
reduces the $W_R$ branching ratios into $jj$, $WZ$, $Wh$, such that a higher production cross section 
becomes needed to fit the data.  Trinification allows to raise the production cross section up to $\tilde g_R = g_R$.
At the same time, the extra decay channels give rise to extra more complicated signals.

In order to keep the discussion simple, let us divide particles into three main categories:
$q$ (generic quarks), $\ell$ (generic leptons), $H$ (Higgses: $h$, $W$, $Z$), all approximatively massless.
Then $q'$ denotes the extra heavy quarks, $\ell'$ the extra heavy leptons, and $H'$ denotes the extra heavy scalars (not eaten by the $W,Z$),
with generic masses $M'$.
The possible $W_R$ decay modes are
\beq W_R\to\left\{\begin{array}{ll}
qq,HH&\hbox{always}\\
qq',\ell \ell', HH' & \hbox{if $M'\circa{<} M_{W_R}$}\\
q'q', \ell'\ell', H'H' & \hbox{if $M'\circa{<} M_{W_R}/2$}
\end{array}\right.\eeq
The extra primed particles, if produced, can decay through gauge interactions as
\beq q'\to q W_R^*,\qquad
\ell' \to \ell W_R^*,\qquad
H'\to H W_R^*\eeq
(where the virtual $W_R^*$ can, in many cases, only materialise into $qq$ or $HH$ light particles)
and through Yukawa interactions as
\beq q'\to q H,\qquad
\ell' \to \ell H,\qquad
H'\to qq, \ell \ell.\eeq
If $W_R\to \ell \ell'$ is kinematically allowed, one obtains final states with two leptons accompanied
by a $jj$ or $WZ$  or $Wh$ pair or by a single $W,Z,h$.
If $\ell'$ is a heavy neutrino, Majorana masses can give rise to same-sign di-leptons, accompanied by the same extra particles listed above.
In particular, for appropriate values of its parameters,  trinification  can also reproduce the $eejj$ excess~\cite{eejj},
as already discussed in~\cite{dibosonWZ} in the context of dedicated models.
Needless to say, all decay widths can be computed in terms of unknown masses and of the mixing angles introduced above.

\section{The di-photon excess at  750 GeV }\label{digamma}
The di-photon excess at 750 GeV~\cite{data} has been tentatively interpreted as a new 750 GeV scalar $S$
with mass $M\approx 750\GeV$
that is produced as $gg\to S$ and decays as $S\to\gamma\gamma$ 
through loops of extra fermions and/or scalars coupled to $S$ by sizeable couplings~\cite{gammagamma,goldrush}.
ATLAS data~\cite{data} possibly indicate a large width $\Gamma_S/M\approx 0.06$, which would make 
this interpretation more difficult. In any case,
the apparently large width can be faked by a multiplet of quasi-degenerate narrow resonances~\cite{gammagamma}.

Trinification has built-in all the ingredients needed to provide such interpretation:
extra scalars $S$ as singlets and/or doublets in the Higgs ${\cal H}$ multiplets;
extra fermions $D'$ and $L'$ that extend the SM chiral fermion content and receive masses only as $y V$;
Yukawa couplings $y$ that must be sizeable enough in order to push the extra fermions above the experimental bounds.


%


\subsection{Trinification predictions}
In order to compute the relevant decay widths of $S$ into $\gamma\gamma$ and $gg$,
we parameterise its linear Lagrangian couplings
to generic fermions $\Q_f$ with Dirac mass $M_f$, 
to generic scalars $\tilde\Q_s$ with mass $M_s$
and to generic vectors $V$ with mass $M_V$ as
\beq S \bar\Q (y_f + y_{5f} \gamma_5) \Q  -  S A_s |\tilde\Q_s|^2 +  g_V M_V S |V_\mu|^2.\eeq
The coupling $g_V$ is normalised such that it coincides with the SM coupling $g_2$, if $S$ is replaced by the SM Higgs $h$
and $V_\mu$ by the SM $W_\mu^\pm$.
Then $S$ acquires the following decay widths at one-loop level~\cite{gammagamma}:
\begin{eqnsystem}{sys:loopf}
\frac{\Gamma(S \to \gamma\gamma)}{M} &=& \frac{\alpha^2}{256\pi^3}\bigg(\bigg|
\sum_f  \frac{y_{5f} M}{M_f} Q_f^2 F_{5f} \bigg|^2 + \nonumber \\
&+&  \bigg|     
\sum_f  \frac{y_f M}{M_f} Q_f^2 F_{f}+
\sum_{s}   \frac{A_s M}{2M_{s}^2} Q_s^2  F_{s}
-\sum_V  \frac{g_V M}{2M_V}  Q_V^2   F_{V} 
\bigg|^2 \bigg)\,,
\label{gamrat}\\
\frac{\Gamma(S \to gg)}{M} &=& \frac{\alpha^2_3}{32\pi^3}\bigg(\bigg|
\sum_f  C_f \frac{y_{5f} M}{M_f}  F_{5f} \bigg|^2 + \bigg|     
\sum_f  C_f \frac{y_f M}{M_f}  F_{f}
\bigg|^2 \bigg),
\end{eqnsystem}
where multiplicities are implicitly included in sums.
The color factors are $C_1=0$ for color singlets and $C_3=1/2$ for color triplets,
and we have taken into account that trinification predicts no other coloured particles.
The loop functions are~\cite{gammagamma}
\beq
\begin{array}{rclll}
F_{s} &=& x[-1 + x f(x)]  &\stackrel{x\to\infty}{=}1/3  & \hbox{with }x = 4M_s^2/M^2 , \\
F_{f} &=& 2 x[1+(1-x) f(x)]  & \stackrel{x\to\infty}{=}4/3  & \hbox{with }x = 4M_f^2/M^2 ,\\
F_{5f} &=& 2x  f(x) &  \stackrel{x\to\infty}{=}2  & \hbox{with }x = 4M_f^2/M^2, \\
F_V &=& 2 + 3 x + 3 x (2-x) f(x)  & \stackrel{x\to\infty}{=}7 & \hbox{with }x = 4M_V^2/M^2,
\label{formfactor}
\end{array}\eeq
where $f(x) = \arctan^2(1/\sqrt{x-1})$.
The $S$ decay width into gluons can be estimated as
\begin{eqnsystem}{sys:GS}
  \frac{\Gamma(S \to gg)}{M} &\approx&  0.6~10^{-4} \left(y_{5D}^2 + \frac49 y_D^2\right)\bigg(\frac{1\TeV}{M_{D'}} \frac{N_{D'}}{3}\bigg)^2  \\
\riga{having assumed, for simplicity, $N_{D'}\le 3$ degenerate copies of $D'$ with mass $M_{D'}\gg M/2$.
This shows  that the $\Gamma(S\to gg)/M\circa{>}10^{-5}$ needed to fit the di-gamma anomaly~\cite{gammagamma} is easily achieved;
furthermore trinification implies that some components of the ${\cal H}$ scalars have sizeable couplings to SM quarks,
allowing for extra production channels of $S$.
The $S$ decay width into photons can be estimated as}\\
\frac{\Gamma(S \to \gamma\gamma)}{M} &\approx&  10^{-6} \bigg[6.7 \bigg(y_{5E} \frac{N_{L'}}{3}\bigg)^2 +  \bigg(1.05 y_E \frac{N_{L'}}{3}+ 1.02\frac{A_s}{M_s} \frac{N_s}{9} \bigg)^2\bigg]
\end{eqnsystem}
having assumed, for simplicity, $N_{L'}\le 3 $ copies of leptons degenerate at $M_{L'} \circa{>}M/2$, 
plus $N_s\le 9$ copies of scalars with charge $Q_s=\pm1$ degenerate at the same mass
and with the same cubic $A_s$.
We neglected the small contribution from heavy down-quarks and from vectors.

\subsection{Comparison with data}
In order to achieve the $\Gamma(S\to\gamma\gamma)/M\circa{>}10^{-6}$ needed to reproduce the $\gamma\gamma$ anomaly~\cite{gammagamma} 
one either needs Yukawa couplings of order unity or larger, or the presence of multiple charged states,
or a large $A_s/M$.

\smallskip

In a context where the theory can be extrapolated up to infinite energy~\cite{TAF,333a},
some scalar quartic couplings can be predicted in terms of squared gauge couplings, times model-dependent order one factors.
This implies that scalars typically have masses comparable to vectors, which must be heavier than about 2 TeV.\footnote{An exception
arises if the breaking of $G_{333}$ proceeds through the Coleman-Weinberg mechanism.
Then the scalar singlet that corresponds to a dilatation of all vacuum expectation values $V$
remains lighter than the other ones by a loop factor.
If this dilaton is the $S$ state, one obtains more neat predictions,
given that vectors and fermions acquire mass only from its vev $\langle S\rangle=V$, such that
$M_V=g_V V/2$, $M_f = y_f  V$, $y_{5f}=0$.
However, in such a case, $\Gamma(S\to\gamma\gamma)$ is somehow too small,
given that such dilaton cannot have a large cubic $A_s$, and that $V\circa{>} \hbox{few TeV}$.
}
One can assume that one of the various scalar doublets contained in ${\cal H}_{1,2,3}$ is accidentally lighter and can be identified with the 125 GeV Higgs boson,
and that one singlet (or doublet) is accidentally light and can be identified with the 750 GeV resonance.
In such a case, the accidental cancellation that makes its mass $M$ smaller than $V$ does not generically suppress,
at the same time, its cubic couplings $A_s $, such that $A_s/M$ can be large
(phenomenologically limited by vacuum decay constraints) enhancing $S\to\gamma\gamma$.
The mixing of $S$ with the physical Higgs $h$ must be small enough.
The two neutral components of a doublet generically receive a small mass splitting $\sim v^2/M \approx 10 \GeV$ and can mimic the large width favoured by ATLAS data~\cite{gammagamma}.
The two components of all electroweak singlets in ${\cal H}$ generically receive mass splittings of order $V$.

\medskip

If the $S\to\gamma\gamma$ loop is dominated by a loop of either lepton doublets or scalar doublets
(which have the same quantum numbers) one predicts the following extra electro-weak decays:
\beq\label{eq:ratio}
\frac{\Gamma(S\to Z\gamma)}{\Gamma(S\to \gamma\gamma)} \approx 0.9< 2,\qquad
\frac{\Gamma(S\to ZZ)}{\Gamma(S\to \gamma\gamma)}\approx 3.5 <6,\qquad
\frac{\Gamma(S\to WW)}{\Gamma(S\to \gamma\gamma)}\approx 10 < 20.
\eeq
These predictions satisfy the experimental bounds collected in~\cite{gammagamma}
and quoted after the $<$ symbols and correspond,
in the language of effective operators of~\cite{gammagamma}, to equal coefficients 
$\Lambda_B=\Lambda_W$, or $\tilde\Lambda_B=\tilde\Lambda_W$.
The extra loop contribution from charged scalar singlets or from right-handed down quarks can reduce the ratios in eq.\eq{ratio}.


\section{Conclusions} \label{concl}
Models based on the trinification gauge group $G_{333}=\SU(3)_L\otimes\SU(3)_R\otimes\SU(3)_c$
explain the quantisation of electric charge~\cite{333}.
Weak scale trinification models provide natural extensions of the Standard Model that can hold up to infinite energy,
with all gauge and Yukawa and quartic couplings flowing to asymptotic freedom~\cite{333a}.

We have shown that these same models, without any extra ingredient,
 can simultaneously explain the $\gamma\gamma$ anomaly at 750 GeV (section~\ref{digamma})
and the diboson anomaly at 1.9 TeV (section~\ref{diboson}).
Run 2 LHC data do not confirm the latter anomaly found in run 1 data, but do not significantly disfavour it
(fig.\fig{fit}a).

The lightest extra vector of trinification is a $W_R^\pm$
with a gauge coupling $g_R\approx 0.444$ compatible with the cross section needed to fit the 1.9 TeV anomaly (fig.\fig{fit}b).
Furthermore, the group structure of trinification implies that it does not decay to SM leptons, in agreement with data.

At the same time, trinification predicts a set of extra fermions (chiral under $G_{333}$ but not under the SM) and of scalars
which can be numerous and light enough to mediate the $gg\to S\to \gamma\gamma$ loops
without giving, at the same time, excessive $gg\to S\to\gamma Z, ZZ, W^+ W^-$.
This interplay between the 2 anomalies implies extra $W_R$ decay channels which give rise to the extra signals discussed in section~\ref{diboson}.

  
 
\footnotesize

\subsubsection*{Acknowledgments}
We thank the ATLAS and CMS colleagues for their beautiful Christmas gift,
hoping that with the restart of data taking on April 1 it will not become a bad joke.
We thank R. Franceschini, M. Redi, J.F. Kamenik
M.McCullogh for useful discussions.
This work was supported by the ERC grant NEO-NAT.

%

\end{document}